%% file: Distributed_Feature_Screening_via_Componentwise_Debiasing.tex
\documentclass[12pt,a4,epsf]{article}
\usepackage{natbib}
\usepackage{graphicx}
\usepackage{latexsym}
\usepackage{epsfig}
\usepackage{bm}
\usepackage{enumerate}
\usepackage{amsmath}
\usepackage{float}
\usepackage{amssymb}
\usepackage{color}
\usepackage{multirow}
\usepackage{amsthm}
\usepackage{mathrsfs}
\usepackage{amsthm}
\usepackage[dvipdfm,CJKbookmarks,bookmarksopen=true,colorlinks=true,
linkcolor=blue,citecolor=blue,pdfstartview=FitH,pdftitle=title,pdfauthor=lixx]{hyperref}
\RequirePackage[mathlines, displaymath]{lineno}      

\input self-define-HL.tex

\newcommand{\be}{\begin{equation}}
\newcommand{\ee}{\end{equation}}
\newcommand{\bea}{\begin{eqnarray}}
\newcommand{\eea}{\end{eqnarray}}
\newcommand{\beas}{\begin{eqnarray*}}
\newcommand{\eeas}{\end{eqnarray*}}
\newtheorem{lemma}{{\bf Lemma}}

\newtheorem{theorem}{{\bf Theorem}}
\newtheorem{proposition}{{\bf Proposition}}

\makeatletter      
\@addtoreset{equation}{section}
\makeatother       

\long \def\@makecation#1#2{ \vskip 10 pt
\setbox\@tempboxa\hbox{#1:#2} \ifdim \wd\@tempboxa >\hsize
\unhbox\@tempboxa\par \else \hbox to\hsize{\hfil\box\@temboxa\hfil}
\fi}

\begin{document}


\title{Distributed Feature Screening via Componentwise Debiasing}
\author{Xingxiang Li$^{a}$,\setcounter{footnote}{-1}\footnote{*Chen Xu is the corresponding
    author. E-mail: cx3@uottawa.ca.}\setcounter{footnote}{-1}\footnote{}\setcounter{footnote}{-1} Runze Li$^{b}$, Zhiming Xia$^{c}$, Chen Xu$^{a*}  \vspace{0.3cm}$ \\
\begin{tabular}{l}
{\small\it $^{a}$Department of Mathematics and Statistics, University of Ottawa, Canada  }\\
{\small\it $^{b}$Department of Statistics and The Methodology
Center, The Pennsylvania State University, USA  }\\
{\small\it $^{c}$School of Mathematics, Northwest University, China }\\
\end{tabular}
}
\date{}   
\maketitle

\begin{quote}
\vspace{-0.6cm}
\begin{abstract}
\noindent Feature screening is a powerful tool in the analysis of high dimensional data. 
When the sample size $N$ and the number of features $p$ are both large, the implementation of classic screening methods can be numerically challenging. In this paper, we propose a distributed screening framework for big data setup. In the spirit of ``divide-and-conquer", the proposed framework expresses a correlation measure as a function of several component parameters, each of which can be distributively estimated using a natural U-statistic from data segments. With the component estimates aggregated, we obtain a final correlation estimate that can be readily used for screening features. This framework enables distributed storage and parallel computing and thus is computationally attractive. Due to the unbiased distributive estimation of the component parameters, the final aggregated estimate achieves a high accuracy that is insensitive to the number of data segments $m$ specified by the problem itself or to be chosen by users. Under mild conditions, we show that the aggregated correlation estimator is as efficient as the classic centralized estimator in terms of the probability convergence bound; the corresponding screening procedure enjoys sure screening property for a wide range of correlation measures. The promising performances of the new method are supported by extensive numerical examples.
\end{abstract}

\noindent {\footnotesize {\it Key words}: Feature screening; Big data; Divide-and-conquer; Componentwise debiasing; Sure screening property.}
\end{quote}

\newpage
\baselineskip=18pt

\section{Introduction}\label{sec1}

With rapid development of data generation and acquisition, massive data with a huge number of features are frequently encountered in many scientific fields. High dimensionality poses simultaneous challenges of computational cost, statistical accuracy, and algorithmic stability for classic statistical methods  (\cite{fan2009ultrahigh}). To facilitate the computing process, one natural strategy is to screen most irrelevant features out before an elaborative analysis. This procedure is referred to as feature screening. With dimensionality reduced from high to low, analytical difficulties are reduced drastically. In the literature, plenty of works have been done in this area; in particular, the correlation-based screening methods have attracted a great deal of attention. These methods conduct screening based on a certain correlation measure between features and the response. Features with weak correlations are treated as irrelevant ones and are to be removed. This type of methods can be conveniently implemented without strong model assumptions (even model-free). Thus, they are commonly used for analyzing high-dimensional data with complex structures. For example, \cite{fan2008sure} proposed a sure independence screening (SIS) based on Pearson correlation. \cite{zhu2011model} proposed a sure independent ranking and screening (SIRS) based on a utility measure that is concerned with the entire conditional distribution of the response given the predictors. \cite{li2012robust} proposed a robust rank correlation screening (RRCS) based on the Kendall $\tau$ rank correlation. \cite{li2012feature} developed a model-free sure independence screening procedure based on the distance correlation (DC-SIS). \cite{wu2015conditional} proposed a distribution function sure independent screening (DF-SIS) approach, which utilizes a measure to test the independence of two variables. \cite{zhou2019model} proposed a robust correlation measure to screen features containing extreme values.

Feature screening has been demonstrated to be an attractive strategy in many applications. Most existing methods are developed under the situation, where the number of features $p$ is large but the sample size $N$ is moderate. However, in modern scientific research, it is increasingly common that data analysts have to deal with big datasets, where $p$ and $N$ are both huge. For example, in modern genome wide genetic studies, millions of SNPs are genotyped on hundreds of thousands participants. In Internet studies, an antivirus software may scan tens of thousands keywords in millions of URLs per minute. When faced with large-$p$-large-$N$ data, the direct implementation of classic screening methods can be numerically inefficient due to storage bottleneck and algorithmic feasibility. For example, for a dataset with $N=p=10,000$, the well-known DC-SIS needs about 60 hours to conduct a full screening on a computer with 3.2 GHz CPU and 32 GB memory. Developing computationally convenient methods for big data screening is therefore desirable in practice.

When a dataset is too huge to be processed on a single computer, it is natural to consider using a ``divide-and-conquer" strategy. In such a strategy, a large problem is first divided into smaller manageable subproblems and the final output is obtained by combining the corresponding sub-outputs. In this spirit, many machine learning and statistical methods have been rebuilt for processing big data (\cite{Zhang2012Comunication}; \cite{Chen2014A}; \cite{Xu2016On},  \cite{battey2018distributed}, \cite{jordan2018communication}). These inspiring works motivate us to explore the feasibility of using this promising strategy for feature screening with big data.

In this paper, we propose a distributed feature screening framework based on aggregated correlation measures, and refer to it as aggregated correlation screening (ACS). In ACS, we express a correlation measure as a function of several component parameters, each of which can be distributively estimated using a natural U-statistic from data segments. With the unbiased component estimates combined together, we obtain an aggregated correlation estimate, which can be readily used for feature screening. In the proposed ACS framework, a massive dataset is split into and processed in $m$ manageable segments, which can be stored in multiple computers and the corresponding local estimations can be done by parallel computing. It thus provides a computationally attractive route for feature screening with large-$p$-large-$N$ data. This framework is also suitable for the setup, where data are naturally stored in different locations. The U-statistic estimation of the component parameters serves as an effective and convenient debasing technique, which ensures the high accuracy of the aggregated correlation estimator and the reliability of the corresponding screening procedure. Under mild conditions, we show that the aggregated correlation estimator is as efficient as the classic centralized estimator in the sense of probabilistic convergence bound. Such a full efficiency is insensitive to the choice of $m$, which may be specified by the problem itself or to be determined by the users. For a wide range of correlation measures, we further show that ACS enjoys the sure screening property without the need of specifying a parametric model (model-free). We demonstrate the computational advantages and promising screening accuracy of ACS in a series of numerical examples.

The rest of this paper is organized as follows. In Section \ref{sec2}, we formulate the research problem and introduce the ACS framework. In Section \ref{sec3}, we investigate the theoretical properties of ACS. In Section \ref{sec4}, we demonstrate the promising performance of ACS by Monte Carlo simulations and a real data example. Concluding remarks are given in Section \ref{sec5} and the proofs of theorems are provided in the Appendix.

\section{Methodology}\label{sec2}

\subsection{Feature screening with big data}\label{sec2-1}
Let $\mathcal{D}=\{(Y_i, \textbf{X}_i)\}_{i=1}^N$ be $N$ independently and identically distributed (i.i.d) copies of $\{Y,\textbf{X}\}$, where $Y$ is a response variable with support $\Phi_y$ and $\textbf{X}=(X_1,...,X_p)^T$ is a $p$-dimensional covariate vector. We are interested in the situation, where $p$ and $N$ are both large. When a dataset is massive and high-dimensional, it is often reasonable to assume that only a handful of covariates (features) are relevant to the response. Let $F(y|\textbf{X})$ be the conditional distribution function of $Y$ given $\textbf{X}$. A feature $X_j$ is considered to be relevant if $F(y|\textbf{X})$ functionally depends on  $X_j$ for some $y\in\Phi_y$. We use $\mathcal{M}$ to denote the index set of the relevant features and define $\mathcal {M}^c=\{1,...,p\}\setminus\mathcal {M}$. The goal of feature screening is to remove most irrelevant features $X_j$s with $j \in \mathcal {M}^c$ before an elaborative analysis.

One commonly used strategy is to first estimate a marginal correlation measure between the response and each feature, and then remove the features with weak correlations. Specifically, let $\omega_j\geq 0$ be a measure of correlation strength between $Y$ and $X_j$. Let $\hat{\omega}_j$ be a centralized estimate of $\omega_j$ based on $\mathcal{D}$. With a pre-specified threshold $\gamma>0$, one may retain the features in
$$
\hat{\mathcal {M}}=\{j: \hat{\omega}_j \geq \gamma, j=1,...,p\},
$$
and remove the others. This classic approach is effective when sample size $N$ is moderate. However, when $N$ and $p$ are both huge, computing $\{\hat{\omega}_j\}_{j=1}^p$ based on the full dataset $\mathcal{D}$ can be numerically costly.

\subsection{Aggregated correlation screening }\label{sec2-2}

Motivated by the recent works in distributed learning, we consider adopting the idea of ``divide-and-conquer" to tackle big data feature screening. Without loss of generality, suppose that the original full dataset $\mathcal{D}$ is equally partitioned into $m$ manageable segments $\{\mathcal{D}_l\}_{l=1}^m$, each of which contains $n=N/m$ observations. Depending on the computational environment, these segments can be distributively stored on and processed by multiple computers or can be sequentially processed by a single computer. Let $\hat{\omega}_{l,j}$ be the local correlation estimate between $X_j$ and $Y$ based on data segment $\mathcal{D}_{l}$. One natural screening strategy is to compute an averaged correlation estimate
\begin{equation}\label{sae}
\bar{\omega}_j=\frac{1}{m}\sum_{l=1}^{m}\hat{\omega}_{l,j}
\end{equation}
for $1\leq j\leq p$ and remove the features with small $\bar{\omega}_j$ values. This approach is referred to as simple average screening (SAS), which is conceptually simple and easy to implement. To facilitate the computing process, using a relatively large number of small segments is often preferred in the analysis. However, when $m$ is large, $\bar{\omega}_j$ may substantially differ from the centralized estimator $\hat{\omega}_j$ due to the cumulated bias inherited from the local estimators. As a result, its screening performance is often unstable in practice, as to be revealed in our numerical studies.

One way to improve SAS is to conduct debiasing on $\hat{\omega}_{l,j}$s before averaging them over. Unfortunately, this is not straightforward for many commonly-used correlation measures that are nonlinear. Our idea is to express a correlation measure $\omega_j$ as a function of several component parameters, and conduct the distributed unbiased estimation of the component parameters. By doing so, we carry out componentwise debasing on original $\hat{\omega}_{l,j}$s in an effective but much easier way. With the unbiased component estimates naturally combined together, we obtain an aggregated correlation estimate that can be readily used for feature screening.

To be more specific, suppose that a correlation measure between $Y$ and $X_j$ can be expressed as
\begin{equation}\label{g_fun}
\omega_j=g(\theta_{j,1},...,\theta_{j,s}),
\end{equation}
where $g$ is a pre-specified function and $\theta_{j,1},...,\theta_{j,s}$ are $s$ component parameters. For a given correlation measure, expression (\ref{g_fun}) may not be unique. We choose the form of $g$ such that the corresponding component parameters can be conveniently estimated with no bias. For the ease of presentation, let $\hat{\theta}_{j,h}(Z_{i_{1}j}, \ldots, Z_{i_{k_{h}}j})$ denote a basis unbiased estimator (kernel) of $\theta_{j,h}$ with the minimal $k_{h}$ i.i.d copies of $Z_j=\{Y,X_j\}$ for $h= 1, \ldots, s$. Without loss of generality, we assume that $\hat{\theta}_{j,h}$ is symmetric such that its value is invariant to the permutation of $\{Z_{i_{1}j}, \ldots, Z_{i_{k_{h}}j}\}$.

Suppose that $\mathcal{D}$ is too big to be processed on a single computer and is equally partitioned into $m$ segments $\{\mathcal{D}_l\}_{l=1}^m$.  We use $\mathcal{S}_l$ to denote the index set of $\{Y,\textbf{X}\}$ copies on $\mathcal{D}_l$. With a pre-specified correlation measure $\omega_{j}$, we propose to distributively screen features in the following framework.
\begin{enumerate}

\item Express $\omega_{j}$ in the form of (\ref{g_fun}) with an appropriate $g$.

\item On each data segment, we estimate $\theta_{j,h}$ by a local U-statistic
\begin{equation}\label{local u}
U^{l}_{j,h} =  \dbinom{n}{k_{h}}^{-1} \sum_{ \{{i_1,...,i_{k_h}}\} \in \mathcal{S}_{l}} \hat{\theta}_{j,h} ( Z_{i_1j},..., Z_{i_{k_h}j}),
\end{equation}
where the summation is over all $\{Z_{i_1j},..., Z_{i_{k_h}j}\}$ combinations chosen from $\mathcal{D}_l$.

\item We compute an aggregated correlation estimate between $Y$ and $X_{j}$ by
\begin{equation}\label{agg u}
\widetilde{\omega}_j=g(\bar{U}_{j,1},...,\bar{U}_{j,s}),
\end{equation}
where $\bar{U}_{j,h} = \frac{1}{m}\sum_{l=1}^{m}U^{l}_{j,h}$ for $h= 1, \ldots, s$.

\item With a user-specified threshold $\gamma >0$, we retain the features in
$$
\widetilde{\mathcal {M}}=\{j: \widetilde{\omega}_j \geq \gamma, j=1,...,p\},
$$
and remove the others.
\end{enumerate}
We name the proposed screening framework as the aggregated correlation screening (ACS). It is seen that step 2 only requires information stored on the data segments, and thus it can be carried out by parallel or sequential processing. This makes ACS computationally suitable for the large-$p$-large-$N$ situation. The use of U-statistics in step 2 helps to further reduce the variances of the local unbiased estimators on $\theta_{j,h}$s and helps to enhance the stability of the method. The computational complexity of (\ref{local u}) is $O(mn^{k_{h}})$, which can be conveniently handled with an appropriate $m$ such that the local sample size $n=N/m$ is moderate. Compared with SAS, ACS screens features based on a non-linear aggregation of unbiased component estimates. This way enables us to substantially reduce the bias of the final correlation estimate with a little sacrifice on the variance. The overall accuracy of the $\omega_{j}$ estimate is therefore improved; this in turn leads to a more reliable screening result in the distritbuted setup.

\subsection{Examples and extension}
\subsubsection{Examples} \label{sec2.3.1}

The proposed ACS framework is suitable for many commonly used correlation measures. We provide a few concrete examples in this subsection. Let $X_{ij}$ denote the $j$th entry of $\textbf{X}_{i}$ defined in Section \ref{sec2-1} for $j= 1, \ldots, p$.

\begin{enumerate}

\item  {\bf Pearson correlation}

Pearson correlation measures the strength of linear relationship between $Y$ and $X_j$. \cite{fan2008sure} utilized it as a feature screening index for the linear model. When Pearson correlation is used in ACS, $\omega_j$ can be expressed in the form of (\ref{g_fun}) by
\[\omega_j=g(\theta_{j,1},...,\theta_{j,5})=\left|\frac{E(X_{j}Y)-E(X_{j})E(Y)}{\sqrt{(EX_{j}^2-E^2(X_{j}))(EY^2-E^2(Y))}}\right|,\]
where $\theta_{j,1}=E(X_{j}Y)$, $\theta_{j,2}=EX_{j}$, $\theta_{j,3}=EY$, $\theta_{j,4}=EX_{j}^2$, and $\theta_{j,5}=EY^2$. In step 2 of ACS, $U^l_{j,h}$ can be computed by (\ref{local u}) with $k_h=1$ and
$$
\hat{\theta}_{j,1}=X_{i_1j}Y_{i_1}, \ \  \hat{\theta}_{j,2}=X_{i_1j},  \  \ \hat{\theta}_{j,3}=Y_{i_1},  \ \  \hat{\theta}_{j,4}=X_{i_1j}^2, \   \ \hat{\theta}_{j,5}=Y^2_{i_1},
$$
for $i_1\in\mathcal{S}_l$. It is seen that $\bar{U}_{j,h}$ in (\ref{agg u}) coincides with classic moment estimates. When the dataset is properly standardized, the expression of $\omega_j$ can be further simplified.

\item {\bf Kendall $\tau$ rank correlation}

Kendall $\tau$  rank correlation measures the ordinal association between $Y$ and $X_j$. It was used in \cite{li2012robust} for feature screening in linear and transformation models. When this correlation measure is used in ACS, $\omega_j$ can be expressed by
\[\omega_j=g(\theta_{j, 1})=\left|E(I(X_{j}<X'_{j})I(Y<Y'))-1/4\right|,\]
where $\{X'_{j},Y'\}$ is an independent copy of $\{X_{j},Y\}$ and $\theta_{j, 1}=E(I(X_{j}<X'_{j})I(Y<Y'))$. In step 2 of ACS, $U^l_{j,1}$ can be computed by (\ref{local u}) with $k_1=2$ and
$$ \hat{\theta}_{j, 1}=\frac{1}{2}\sum_{(i_1,i_2)} I(X_{i_1j}<X_{i_2j})I(Y_{i_1}<Y_{i_2}),$$
where $\{i_1,i_2\}\in \mathcal{S}_l$ and the summation is over all permutations of $(i_1,i_2)$.

\item {\bf SIRS correlation}

SIRS correlation can be used to detect nonlinear relationship between $Y$ and $X_j$. It was proposed by \cite{zhu2011model} for feature screening in parametric and semiparametric models. When this correlation is used in ACS, $\omega_j$ can be expressed by
\[\omega_j=\theta_{j,1}=E_{Y'}\{E^2(X_jI(Y<Y'))\},\]
where $Y'$ is an independent copy of $Y$ and feature $X_j$ is assumed to have zero mean and unit variance.
In step 2 of ACS, $U_{j,1}^{l}$ can be computed by (\ref{local u}) with $k_1=3$ and
\begin{eqnarray*}
\hat{\theta}_{j,1}=\frac{1}{6}\sum_{(i_1,i_2,i_3)}X_{i_1j}X_{i_2j}I(Y_{i_1}<Y_{i_3})I(Y_{i_2}<Y_{i_3}),
\end{eqnarray*}
where $\{i_1,i_2,i_3\}\in \mathcal{S}_l$ and the summation is over all permutations of $(i_1,i_2,i_3)$.

\item {\bf Distance correlation}

Distance correlation (DC) can be used to measure the dependence between $Y$ and $X_j$.  \cite{li2012feature} utilized it as a model-free screening index. When DC is used in ACS, $\omega_j$ can be expressed by
\[\omega_j=g(\theta_{j,1},...,\theta_{j,8})=\frac{\theta_{j,1}+\theta_{j,2}\cdot\theta_{j,3}-2\theta_{j,4}}{\sqrt{(\theta_{j,5}+\theta_{j,2}^2-2\theta_{j,6})(\theta_{j,7}+\theta_{j,3}^2-2\theta_{j,8})}}\] with
\begin{eqnarray*}
&&\theta_{j,1}=E\{|Y-Y'|\cdot|X_j-X'_j|\}, \nonumber\\
&& \theta_{j,2}=E\{|{Y}-{{Y}'}|\}, \quad\theta_{j,3}=E\{|{X}_j-{X}'_j|\}, \nonumber\\
&& \theta_{j,4}=E\{E(|{Y}-{{Y}'}|\ |\ {Y})E(|{X}_j-{X}'_j| \ | \ {X}_j)\},\nonumber\\
&&\theta_{j,5}=E\{|{Y}-{{Y}'}|^2\}, \quad \theta_{j,6}=E\{E^2(|{Y}-{Y}'| \ | \ {Y})\}, \nonumber\\
&&\theta_{j,7}=E\{|{X}_j-{X}'_j|^2\}, \quad \theta_{j,8}=E\{E^2(|{X}_j-{X}'_j| \ | \ {X}_j)\},
\end{eqnarray*}
where $(Y',X'_j)$ is an independent copy of $(Y,X_j)$. In step 2 of ACS, $U^l_{j,1}$, $U^l_{j,4}$ can be computed by (\ref{local u}) with $k_{1}=2$, $k_{4}=3$, and
\begin{eqnarray} \label{eq-dc-1}
\hat{\theta}_{j,1}&=&\frac{1}{2}\sum_{(i_1,i_2)}|{Y}_{i_1}-{Y}_{i_2}|\cdot|{X}_{i_1j}-{X}_{i_2j}|, \\ \label{eq-dc-2}
\hat{\theta}_{j,4}&=&\frac{1}{6}\sum_{(i_1,i_2,i_3)} |{Y}_{i_1}-{Y}_{i_3}|\cdot|{X}_{i_2j}-{X}_{i_3j}|.
\end{eqnarray}
The expression of $\hat{\theta}_{j,h}$ for $h=2,3,5,7$ is similar to (\ref{eq-dc-1}); the expression of $\hat{\theta}_{j,h}$ for $h=6, 8$ is similar to (\ref{eq-dc-2}).
\end{enumerate}

\noindent
{\bf Remark:} When Pearson correlation is used, the aggregated estimator $\widetilde{\omega}_j$ in (\ref{agg u}) coincides with the centralized estimator $\hat{\omega}_j$; the proposed ACS leads to the same screening result of the classic SIS. For the correlations in Examples 2-4, the computational cost of $\widetilde{\omega}_j$ is substantially lower than that of $\hat{\omega}_j$. For Kendall $\tau$ correlation, ACS reduces the computational complexity in correlation estimation from $O(N^2)$ down to $O(N^2/m)$. When the data segments are parallel processed and the communication cost is negligible, the computational time of ACS decreases drastically when $m$ increases.

The idea of componentwise debiasing in ACS provides a viable and effective route to estimate $\omega_{j}$ in a distributed manner. For commonly-used correlation measures, form (\ref{g_fun}) can be naturally constructed. The simplicity and compatibility of ACS make it a user-friendly approach in practice.

\subsubsection{Extension}\label{sec2.3.2}
When data partition is manually done, one may further improve the stability of ACS with multiple partitions. Specifically, suppose that we repeat the random data partition $R$ times. For each partition, we conduct unbiased estimation of component parameters based on (\ref{local u}). We then carry out (\ref{agg u}) with $\bar{U}_{j,h}$ replaced by
\[
\breve{U}_{j,h}^R=\frac{1}{R}\sum_{r=1}^R \bar{U}_{j,h}^r,
\]
where $\bar{U}_{j,h}^r$ denotes the mean U-statistic for the $r$th partition. By averaging over $R$ partitions, the variability of $\widetilde{\omega}_{j}$ is further reduced; this leads to a reinforced ACS that is more reliable for feature screening.

\section{Theoretical Analysis} \label{sec3}
We now provide some theoretical justification of using ACS. Apparently, the screening performance of ACS relies on the accuracy of the aggregated correlation estimator $\widetilde{\omega}_j$ (\ref{agg u}). We show that
$\widetilde{\omega}_j$ is an effective and efficient tool to estimate $\omega_j$; this serves as a theoretical foundation of ACS. Our theoretical investigation is based on the following technical conditions.

\vspace{0.2cm}
\noindent
\begin{enumerate}[C1]

\item
There exists a constant $\kappa_0>0$ such that, for any $0\leq\kappa\leq\kappa_0$, $E\{\exp(\kappa\hat{\theta}_{j,h})\}<\infty$ for all $h=1,...,s$, $j=1,...,p$.

\item
In (\ref{g_fun}), $g(\cdot)$ is formed by finite operations of addition, subtraction, multiplication,  division, absolutization, and square root, where the denominator in division is non-zero and the square root is taken over a positive quantity.

\item  There exist two constants $c>0$ and $0<\tau<1/2$ such that $\min \limits_{j \in \mathcal {M}} \omega_j \geq 2cN^{-\tau}.$
\end{enumerate}

\vspace{0.3cm}
\noindent
Condition C1 requires that $\hat{\theta}_{j,h}$ has a regular distribution, such that its moment generating function exists on $[0,\kappa_0]$. This is a mild condition for many correlation measures. For example, when Kendall $\tau$ correlation is used with ACS, $\hat{\theta}_{j,h}$ is bounded and thus C1 is naturally satisfied; when SIRS is used with ACS, C1 is implied if $E\{\exp( \xi X_{j}^{2}) \} < \infty$ for some $\xi > 0$ and $1 \leq  j \leq p$. Condition C2 is applicable to a variety of commonly used correlation measures, including the ones discussed in Section \ref{sec2.3.1}. We conjecture that ACS would still be effective with a more complicated $g(\cdot)$. However, the corresponding theoretical justification is likely to be lengthy. Here, we aim to provide some theoretical understanding of the proposed screening framework and do not intend to make this condition weakest possible. Condition C3 requires that the marginal correlation between any relevant feature and the response should not be too small. This is a natural feature identifiability requirement, which has been widely used in the literature; see, for example, Condition 3 of \cite{fan2008sure}, Condition 2 of \cite{li2012feature}, and Condition 6 of \cite{wu2015conditional}.

With the conditions above, we derive a probability inequality for $\bar{U}_{j,h}$ in the following proposition; it serves as a prerequisite for the effectiveness of $\widetilde{\omega}_j$.

\vspace{0.5cm}
\noindent
\begin{proposition}\label{prop1}
Suppose Condition C1 is satisfied and $\varepsilon\in (0,\delta_0]$ with an arbitrarily large $\delta_0>0$. There exists a sufficiently small $c_0>0$ such that
$$
P(|\bar{U}_{j,h}-\theta_{j,h})|\geq \varepsilon)\leq 2(1-c_0\varepsilon^2/2)^{m \lfloor n/k_h\rfloor},
$$
for $j= 1, \ldots, p$ and $h= 1, \ldots, s$, where $\lfloor n/k_h\rfloor$ denotes the largest integer no larger than $n/k_h$.
\end{proposition}

\vspace{0.5cm}\noindent
Proposition \ref{prop1} can be viewed as an generalization of the classic Berk's inequality for the distributed setup with $m > 1$ (\cite{Robert1966Limiting}). It also echoes Theorem 2 of \cite{lin2010fast} in a non-asymptotical sense. Proposition \ref{prop1} implies that the component parameters can be effectively estimated by summarizing the corresponding local U-statistics from data segments. With Proposition \ref{prop1}, we show the effectiveness of $\widetilde{\omega}_j$ in the following theorem.

\vspace{0.5cm}
\noindent
\begin{theorem} \label{Thm1}
Suppose that Conditions C1-C3 are satisfied and $k= \max\{k_{h}, h= 1, \ldots, s\} \leq n$. There exists a constant $\eta>0$ such that
\begin{eqnarray*}
P\left(\max \limits_{1 \leq j \leq p} |\widetilde{\omega}_j-\omega_j|\geq cN^{-\tau} \right)\leq  \eta p (1-N^{-2\tau}/ \eta)^{m\lfloor n/k \rfloor}.
\end{eqnarray*}
\end{theorem}

\vspace{0.5cm}
\noindent Note that $k$ is a constant depending on the choice of $\omega_{j}$ and $g(\cdot)$; thus, ${m\lfloor n/k \rfloor}$ is in the same order of $N$. Theorem \ref{Thm1} implies that the aggregated correlation estimators are uniformly consistent even when $p$ grows exponentially with $N^{\alpha}$ for some $0<\alpha<1$. 
In the literature, it has been shown that the centralized estimator achieves convergence bound $|\hat{\omega}_j-\omega_j| = O_{p}(N^{-\tau})$ for $0<\tau<1/2$ (\cite{li2012robust}, \cite{li2012feature}, \cite{cui2015model}, \cite{wu2015conditional}). Theorem \ref{Thm1} indicates that $\widetilde{\omega}_j$ works as efficiently as the centralized estimator $\hat{\omega}_j$. Benefited from the unbiased estimation of the component parameters, the high efficiency $\widetilde{\omega}_j$ does not depend on the choice of $m$; this leads to a reliable feature screening. We justify the proposed ACS framework using the following theorem.

\vspace{0.5cm}
\begin{theorem} \label{Thm2}
Under Conditions C1-C3, if $k\leq n$ and $\gamma=cN^{-\tau}$, then there exists a constant $\eta>0$ such that
\begin{eqnarray*}
P\{\mathcal{M}\subseteq\widetilde{\mathcal {M}} \}\geq 1- \eta d(1-N^{-2\tau}/\eta)^{m\lfloor n/k \rfloor},
\end{eqnarray*}
where $d$ is the cardinality of $\mathcal {M}$.
\end{theorem}

\vspace{0.5cm}
\noindent Theorem \ref{Thm2} shows that the proposed ACS enjoys sure screening property in the sense of \cite{fan2008sure}, even when the number of relevant features $d$ is diverging. That is, when $N$ is large, ACS removes most irrelevant features and retains all relevant features with an overwhelming probability. It is a desired property for a good feature screening method.   Note that the requirement $n=N/m \geq k$ is very mild in general; for many correlation measures, it can be naturally satisfied with a liberal choice of $m=O(N)$. Compared with SAS, ACS is less sensitive to the choice of $m$; this makes it a flexible and reliable approach. Our empirical experiences show that a small $m$ may help to improve the practical screening accuracy of ACS. However, an overly small $m$ often leads to a high computational cost. In applications, one good strategy is to choose the smallest $m$ for ACS within the computational budget.

\section{Numerical Studies}\label{sec4}

We assess the finite sample performance of ACS via simulations and a real data example. In particular, we compare ACS with the naive SAS in terms of the screening accuracy and stability. All numerical experiments are conducted using software MATLAB on Windows computers with 3.2 GHz CPUs and 32 GB memory.


\subsection{Simulations}\label{sec3-1}

\begin{figure}[t!]
\setlength{\parindent}{-2.5em} \centering
\scalebox{0.85}{\includegraphics{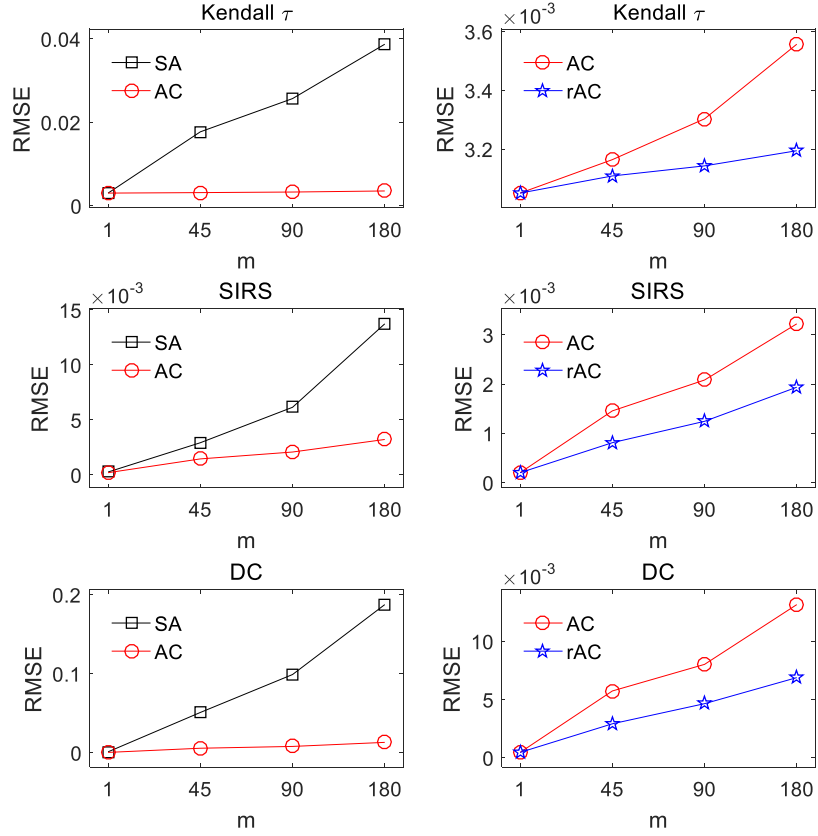}}
\caption{The accuracy of distributed correlation estimators: SA, AC, and rAC stand for $\bar{\omega}_j$, $\widetilde{\omega}_j$, and reinforced $\widetilde{\omega}_j$ respectively. }\label{fig1}
\end{figure}

Apparently, an effective screening relies on the accurate estimates of the correlation strength $\omega_{j}$. Our first experiment is to check whether the proposed aggregated correlation (AC) measure $\widetilde{\omega}_j$ in (\ref{agg u}) is an effective estimator of $\omega_{j}$. To this end, we generate $N=2700$ independent copies from $(Y, X)$, where $Y$ and $X$ are two independent random variables following $N(0, 1)$. Due to independence, the Kendall $\tau$ correlation,  SIRS, and DC between $Y$ and $X$ are all zero. We randomly split the data into $m= 45, 90, 180$ equal-sized segments and use $\widetilde{\omega}_j$ specified in Section \ref{sec2.3.1} (with $j=1$) to estimate the three aforementioned correlations between $Y$ and $X$. We repeat the procedure $T=500$ times and measure the accuracy of $\widetilde{\omega}_j$ by root-mean-squared error (RMSE). Specifically, let $\widetilde{\omega}_j (t)$ denote the value of  $\widetilde{\omega}_j$ for the $t$th repetition. RMSE is computed by \[\mbox{RMSE}(\widetilde{\omega}_j) =  \left[\frac{1}{T}\sum_{t=1}^{T}(\widetilde{\omega}_j(t) )^2 \right]^{1/2} .\]
For comparison, we report the corresponding RMSEs of the simple averaging (SA) estimators $\bar{\omega}_j$ defined in (\ref{sae}) under the same $m$ setup. Moreover, we check the performance of the reinforced $\widetilde{\omega}_j$ (rAC) using the multiple partition strategy with $R=3$ as discussed in Section \ref{sec2.3.2}. As a benchmark, we also report the RMSEs of the centralized estimators with $m=1$. The results are summarized in Figure \ref{fig1} with the corresponding computational time (in seconds) given in Table \ref{Table1}.

For all the three tested correlations, we see that both $\widetilde{\omega}_j$ and $\bar{\omega}_j$ work well when $m$ is small. As $m$ increases, $\bar{\omega}_j$ becomes less accurate. As discussed, this is mainly due to the non-negligible biases of the segmental estimates. In comparison, $\widetilde{\omega}_j$ conducts componentwise debiasing and leads to a high estimation accuracy over a wide range of $m$. Compared with the centralized estimators ($m=1$ case), the distributed estimators $\widetilde{\omega}_j$ and $\bar{\omega}_j$ are computationally more attractive, in particular when $m$ is large. As expected, the reinforced aggregated estimators help to further improve the estimation accuracy of $\widetilde{\omega}_j$ at a higher computational cost.

\begin{table}[t!]
\footnotesize
\caption{Computational time of distributed correlation estimators (in seconds) }\label{Table1}
\begin{tabular*}{\textwidth}{@{\extracolsep{\fill}}cccccc}
\hline
\hline
Correlation      &   Estimator     &  $m=1$               & $m=45$              &  $m=90$              &  $m=180$             \\
\hline
                      & SA         &  $3.4\cdot 10^{-1}$  & $1.4\cdot 10^{-4}$  &  $4.3\cdot 10^{-5}$  &  $2.2\cdot 10^{-5}$  \\
Kendall $\tau$        & AC         &  $3.4\cdot 10^{-1}$  & $1.5\cdot 10^{-4}$  &  $4.7\cdot 10^{-5}$  &  $2.6\cdot 10^{-5}$  \\
                      & rAC        &  $--$                & $3.7\cdot 10^{-4}$  &  $1.2\cdot 10^{-4}$  &  $5.5\cdot 10^{-5}$  \\
\hline
                      & SA         &  $1.7\cdot 10^{-1}$  & $1.1\cdot 10^{-4}$  &  $5.0\cdot 10^{-5}$  &  $4.0\cdot 10^{-5}$  \\
SIRS                  & AC         &  $1.7\cdot 10^{-1}$  & $7.5\cdot 10^{-5}$  &  $2.5\cdot 10^{-5}$  &  $1.4\cdot 10^{-5}$  \\
                      & rAC        &  $--$                & $2.0\cdot 10^{-4}$  &  $7.0\cdot 10^{-5}$  &  $3.9\cdot 10^{-5}$  \\
\hline
                      & SA         &  $8.2\cdot 10^{-1}$  & $1.4\cdot 10^{-4}$  &  $4.7\cdot 10^{-5}$  &  $3.3\cdot 10^{-5}$  \\
DC                    & AC         &  $7.8\cdot 10^{-1}$  & $1.3\cdot 10^{-4}$  &  $4.0\cdot 10^{-5}$  &  $2.7\cdot 10^{-5}$  \\
                      & rAC        &  $--$                & $3.6\cdot 10^{-4}$  &  $1.2\cdot 10^{-4}$  &  $9.3\cdot 10^{-5}$  \\
\hline
\hline
\end{tabular*}
\end{table}

The promising performance of $\widetilde{\omega}_j$ encourages us to further check whether the associated screening procedure ACS also works well. To this end, we generate $N$ independent copies of $\textbf{X}=(X_{1}, \ldots, X_{p})$ from a multivariate normal distribution with zero mean. The corresponding response $Y$ is generated based on the following models.
\begin{eqnarray*}
&(\textnormal{a})&\ Y=\beta_1X_1+\beta_2X_2+...+\beta_8X_{8}+\varepsilon,  \nonumber\\
&(\textnormal{b})&\ Y=\beta_1X_1+\beta_2X_4+\beta_3X_7+\beta_4X_{10}+\varepsilon, \nonumber\\
&(\textnormal{c})&\ Y=\exp(\beta_1X_1+\beta_2X_4+\beta_3X_7+\beta_4X_{10}+\varepsilon), \nonumber\\
&(\textnormal{d})&\ Y=\beta_1X_1+\beta_2X_4+\exp(|\beta_3|X_7+|\beta_4|X_{10})+\varepsilon,  \nonumber\\
&(\textnormal{e})&\ Y=\beta_1X_1+\beta_2X_4^2+\beta_3I(X_{7}>0)+\beta_4|X_{10}|+\varepsilon, \nonumber\\
&(\textnormal{f})&\ Y=2\beta_1X_1X_2+2\beta_2I(X_{12}>0)+3\beta_3X_{22}+\varepsilon,
\end{eqnarray*}
where $\varepsilon\sim N(0,1)$ is a noise term.
Models (a) and (b) are two linear cases with different model sparsity and covariance structures. Models (c) and (d) are transformation model and multiple-index model, which are adopted from \cite{li2012robust} and \cite{zhu2011model}) respectively. Models (e) and (f) are addictive model and interactive model, both of which were discussed in \cite{li2012feature}. In Model (a), $\mbox{cov}(\textbf{X})$ is set to be an identity matrix, while in Models (b)-(f) we set $\textnormal{cov}(X_j,X_r)=0.5^{|j-r|}$ for $j,r \in \{1, \ldots, p\}$ such that the features have an autoregressive correlation. In Models (a)-(f), the values of model coefficients are generated by $(-1)^{W}(2+|V|)$, where $W \sim \mbox{Bernoulli}(0.6)$ and $V\sim N(0,1)$.

We apply the proposed ACS on these simulated datasets for feature screening. In each case, we split the data into $m$ segments and assess the performance of ACS based on Pearson, Kendall $\tau$, SIRS, and DC correlations as discussed in Section \ref{sec2.3.1}. For each correlation scenario, we set the corresponding screening threshold by
\begin{eqnarray}\label{gamma}
\gamma=\rho \cdot \min_{j\in \mathcal{M}} \hat{\omega}_j,
\end{eqnarray}
where $\hat{\omega}_j$ is the centralized estimator of that correlation and $\rho=0.8, 0.6$ is a scale parameter. The choice of $\gamma$ in (\ref{gamma}) guarantees that all relevant features will be retained by the classic screening method based on $\hat{\omega}_j$; it purely serves for the purpose of evaluating the proposed ACS. In practice, a proper $\gamma$ is usually determined by users based on their research goals as well as the prior information about their data.

\begin{table}[t]
\footnotesize
\caption[aa]{Simulation results for Model (a) with $N=1500$, $p=1500$, $\|\mathcal{M}\|_{0}=8$.  The two values $a,b$ in the same column correspond to $\rho= 0.8, 0.6$ cases. }\label{Table2}
\vspace{0.2cm}
\centering
\begin{tabular*}{\textwidth}{@{\extracolsep{\fill}}ccc|ccccc|cc}
\hline
\hline
$m$	   &  Correlation & Method     & SSR          &   MS      & Std(MS)   &  PSR          &  FDR       & $\mbox{Time}^n$   & $\mbox{Time}^N$\\
\hline
15     &  Pearson        & SAS     & 1.0, 1.0     & 8, 8      & 0, 7      &  1.0, 1.0     &  0.0, 0.0  &  0.001            &0.020 \\
       &                 & ACS     & 1.0, 1.0     & 8, 8      & 0, 0      &  1.0, 1.0     &  0.0, 0.0  &  0.001            &\\
       &  Kendall $\tau$ & SAS     & 1.0, 1.0     & 8, 8      & 0, 5      &  1.0, 1.0     &  0.0, 0.0  &  0.244            &81.42 \\
       &                 & ACS     & 1.0, 1.0     & 8, 8      & 0, 0      &  1.0, 1.0     &  0.0, 0.0  &  0.244            &\\
       &                 & rACS    & 1.0, 1.0     & 8, 8      & 0, 0      &  1.0, 1.0     &  0.0, 0.0  &  0.732            &\\
       &  SIRS           & SAS     & 1.0, 1.0     & 8, 9      & 3, 52     &  1.0, 1.0     &  0.0, .11  &  0.003            &0.150  \\
       &                 & ACS     & .72, .94     & 14, 33    & 13, 26    &  1.0, 1.0     &  .43, .76  &  0.003            &\\
       &                 & rACS    & .94, .99     & 8, 8      & 1, 3      &  .99, .99     &  0.0, 0.0  &  0.010            &\\
       &  DC             & SAS     & 1.0, 1.0     & 20, 716   & 382, 572  &  1.0, 1.0     &  .59, .99  &  0.306            &177.9\\
       &                 & ACS     & .97, 1.0     & 8, 8      & 0, 1      &  1.0, 1.0     &  0.0, 0.0  &  0.306            &\\
       &                 & rACS    & 1.0, 1.0     & 8, 8      & 0, 0      &  1.0, 1.0     &  0.0, 0.0  &  0.910            &\\
\hline
30     &  Pearson        & SAS     & 1.0, 1.0     & 8, 8      & 5, 148    &  1.0, 1.0     &  0.0, 0.0  &  0.001            &0.020 \\
       &                 & ACS     & 1.0, 1.0     & 8, 8      & 0, 0      &  1.0, 1.0     &  0.0, 0.0  &  0.001            &\\
       &  Kendall $\tau$ & SAS     & 1.0, 1.0     & 8, 56     & 3, 178    &  1.0, 1.0     &  0.0, .86  &  0.071            &81.42 \\
       &                 & ACS     & 1.0, 1.0     & 8, 8      & 0, 0      &  1.0, 1.0     &  0.0, 0.0  &  0.071            &\\
       &                 & rACS    & 1.0, 1.0     & 8, 8      & 0, 0      &  1.0, 1.0     &  0.0, 0.0  &  0.211            &\\
       &  SIRS           & SAS     & 1.0, 1.0     & 61, 779   & 385, 510  &  1.0, 1.0     &  .87, .99  &  0.001            &0.150  \\
       &                 & ACS     & .60, .76     & 39, 89    & 31, 47    &  1.0, 1.0     & .80, .91   &  0.001            &\\
       &                 & rACS    & .85, .98     & 8, 14     & 4, 14     &  1.0, 1.0     &  0.0, .41  &  0.003            &\\
       &  DC             & SAS     & 1.0, 1.0     & 1500, 1500& 147, 0   &  1.0, 1.0      & .99, .99   &  0.087            &177.9  \\
       &                 & ACS     & .90, .99     & 8, 8      & 0, 3      &  1.0, 1.0     & 0.0, 0.0   &  0.087            &\\
       &                 & rACS    & .98, .99     & 8, 8      & 0, 0     &  1.0, 1.0      & 0.0, 0.0   &  0.255            &\\
              \hline
              \hline
\end{tabular*}
\end{table}

We evaluate the performance of ACS in terms of successful screening rate (SSR), screened model size (MS), positive selection rate (PSR), false discovery rate (FDR), Specifically, let $\hat{\mathcal {M}}(t)$ denote the index set of the features retained after screening based on the $t$-th repetition. The aforementioned four indices are calculated as follows.
\begin{eqnarray*}
\textnormal{SSR}=\frac{1}{T}\sum\limits_{t=1}^{T}  I_{\{\mathcal {M} \subset \hat{\mathcal {M}}({t}) \}}, \quad \textnormal{MS} = \left\lfloor \|\hat{\mathcal {M}}({t})\|_0 \right \rfloor_{med}, \\
\textnormal{PSR}=  \left\lfloor \frac{\|\mathcal {M} \cap  \hat{\mathcal {M}}({t})\|_0}{\|\mathcal {M}\|_0} \right \rfloor_{med}, \quad \textnormal{FDR}= \left\lfloor \frac{\| \hat{\mathcal {M}}({t})-\mathcal {M}\|_0}{\|\hat{\mathcal {M}}({t})\|_0} \right \rfloor_{med},
\end{eqnarray*}
where $I_{\{\cdot\}}$ is an indicator function, $\lfloor \cdot \rfloor_{med}$ denotes the median of a series of values, and $\|\cdot\|_{0}$ denotes the number of elements in a set. For comparison, we report as well the screening outcomes of SAS, which is based on the simple averaging estimators (\ref{sae}). To check the improving strategy in Section \ref{sec2.3.2}, we further run the reinforced ACS (rACS) with $R=3$ for the data generated from Model (a).
We summarize the simulation results in Tables \ref{Table2}-\ref{Table4} based on $T=100$ repetitions. For Models (c)-(f), we only exhibit the selected results due to the page limit. In the tables, $\mbox{Time}^{n}$ and $\mbox{Time}^{N}$ report the averaged computational time (in seconds) respectively for a distributed screening and the corresponding classic screening based on centralized correlation estimators. The two values in the same column correspond to the two setups of $\rho$ in (\ref{gamma}). $\mbox{Std}(\textnormal{MS})$ reports the sample standard deviation of $\|\hat{\mathcal {M}}({t})\|_0$s, which measures the screening precision.

\begin{table}[t]
\footnotesize
\caption{Simulation results for Model (b) with $N=1200$, $p=1500$, $\|\mathcal{M}\|_{0}=4$. }
\setlength{\tabcolsep}{2.3mm}
\label{Table3}
\centering
\begin{tabular*}{\textwidth}{@{\extracolsep{\fill}}ccc|ccccc|cc}
\hline
\hline	
$m$  &   Correlation       & Method    & SSR          & MS          & Std(MS)  &  PSR          &  FDR       &  $\mbox{Time}^n$ & $\mbox{Time}^N$\\
\hline
20   &	 Pearson           & SAS       & 1.0, 1.0     & 7, 9        & 19, 110  &  1.0, 1.0     &  .43, .56  &  0.001  & 0.022\\
     &                     & ACS       & 1.0, 1.0     & 7, 8        & 2, 2     &  1.0, 1.0     &  .43, .50  &  0.001  & \\
     &   Kendall $\tau$    & SAS       & 1.0, 1.0     & 7, 9        & 19, 115  &  1.0, 1.0     &  .43, .56  &  0.100  & 51.20   \\
     &                     & ACS       & 1.0, 1.0     & 6, 8        & 2, 1     &  1.0, 1.0     &  .33, .50  &  0.100  & \\
     &   SIRS              & SAS       & 1.0, 1.0     & 7, 9        & 180, 315 &  1.0, 1.0     &  .43, .56  &  0.002  & 0.111 \\
     &                     & ACS       & .83, .95     & 6, 10       & 39, 56   &  1.0, 1.0     &  .43, .60  &  0.001  & \\
     &   DC                & SAS       & 1.0, 1.0     & 9, 14       & 546, 627 &  1.0, 1.0     &  .56, .71  &  0.108  & 105.3  \\
     &                     & ACS       & .98, 1.0     & 6, 7        & 3, 9     &  1.0, 1.0     &  .33, .43  &  0.108  & \\
\hline
40   &   Pearson           & SAS       & 1.0, 1.0     & 7, 10       & 135, 287 &  1.0, 1.0     &  .43, .60  &  0.001  & 0.022\\
     &                     & ACS       & 1.0, 1.0     & 7, 8        & 2, 2     &  1.0, 1.0     &  .43, .50  &  0.001  & \\
     &   Kendall $\tau$    & SAS       & 1.0, 1.0     & 7, 11       & 150, 429 &  1.0, 1.0     &  .43, .64  &  0.033  & 51.20   \\
     &                     & ACS       & 1.0, 1.0     & 7, 8        & 2, 1     &  1.0, 1.0     &  .43, .50  &  0.033  & \\
     &   SIRS              & SAS       & 1.0, 1.0     & 10, 151     & 567, 633 &  1.0, 1.0     &  .60, .97  &  0.001  & 0.111  \\
     &                     & ACS       & .81, .92     & 9, 26       & 65, 89   &  1.0, 1.0     &  .53, .84  &  0.001  & \\
     &   DC                & SAS       & 1.0, 1.0     & 1425, 1500  & 667, 449 &  1.0, 1.0     &  .99, .99  &  0.038  & 105.3  \\
     &                     & ACS       & .95, 1.0     & 6, 7        & 10, 18   &  1.0, 1.0     &  .33, .43  &  0.038  & \\
\hline
\hline
\end{tabular*}
\end{table}

\begin{table}[h!]
\footnotesize
\caption{Simulation results for Models (c)-(f). }\label{Table4}
\centering
\begin{tabular*}{\textwidth}{@{\extracolsep{\fill}}p{1.8mm}cc|ccccc|cc}
\hline
\hline
$m$&  Correlation    &  Method & SSR       &  MS          & Std(MS)    &  PSR          &  FDR       & $\mbox{Time}^n$ & $\mbox{Time}^N$\\
\hline
\multicolumn{10}{c}{Model (c), $N=2400$, $p=2500$, $\|\mathcal{M}\|_{0}=4$} \\
\hline
40 &  Pearson        &  SAS    & 1.0, 1.0  & 1379, 2067   & 700, 539   &  1.0, 1.0     &  .99, .99  &  0.002 & 0.054 \\
   &                 &  ACS    & 1.0, 1.0  & 509, 853     & 593, 584   &  1.0, 1.0     &  .99, .99  &  0.002 & \\
   &  Kendall $\tau$ &  SAS    & 1.0, 1.0  & 7, 9         & 2, 60      &  1.0, 1.0     &  .38, .56  &  0.164 & 357.3 \\
   &                 &  ACS    & 1.0, 1.0  & 7, 9         & 2, 2       &  1.0, 1.0     &  .38, .53  &  0.164 & \\
\hline
80 &  Pearson        &  SAS    & 1.0, 1.0  & 1613, 2202   & 687, 515   &   1.0, 1.0    &  .99, .99  &  0.002 &0.054 \\
   &                 &  ACS    & 1.0, 1.0  & 509, 853     & 593, 584   &   1.0, 1.0    &  .99, .99  &  0.002 & \\
   &  Kendall $\tau$ &  SAS    & 1.0, 1.0  & 7, 10        & 135, 703   &  1.0, 1.0     &  .43, .60  &  0.055 &357.3 \\
   &                 &  ACS    & 1.0, 1.0  & 6, 8         & 2, 2       &  1.0, 1.0     &  .33, .50  &  0.055 & \\
\hline
\multicolumn{10}{c}{Model (d), $N=3600$, $p=3600$, $\|\mathcal{M}\|_{0}=4$} \\
\hline
50 &  Pearson        &  SAS    & 1.0, 1.0  & 3600, 3600   & 433, 305   &  1.0, 1.0     &  .99, .99  &  0.004  &0.099 \\
   &                 &  ACS    & 1.0, 1.0  & 2723, 2940   & 720, 574   &  1.0, 1.0     &  .99, .99  &  0.004  & \\
   &  SIRS           &  SAS    & 1.0, 1.0  & 8, 10        & 72, 470    &  1.0, 1.0     &  .50, .60  &  0.006  &1.568 \\
   &                 &  ACS    & .98, .99  & 7, 8         & 8, 23      &  1.0, 1.0     &  .43, .50  &  0.006  & \\
\hline
100&  Pearson        &  SAS    & 1.0, 1.0  & 3600, 3600   & 397, 277   &  1.0, 1.0     &  .99, .99  &  0.003  &0.099 \\
   &                 &  ACS    & 1.0, 1.0  & 2723, 2940   & 720, 574   &  1.0, 1.0     &  .99, .99  &  0.003  &\\
   &  SIRS           &  SAS    & 1.0, 1.0  & 11, 22       & 1094, 1468 &  1.0, 1.0     &  .64, .81  &  0.003  &1.568  \\
   &                 &  ACS    & .95, .99  & 7, 10        & 31, 68     &  1.0, 1.0     &  .43, .60  &  0.003  & \\
\hline
\multicolumn{10}{c}{Model (e), $N=4800$, $p=4800$, $\|\mathcal{M}\|_{0}=4$} \\
\hline
60 &  Pearson        &  SAS    & 1.0, 1.0  & 4800, 4800   & 0, 0       &  1.0, 1.0     &  .99, .99  &  0.006  &0.172 \\
   &                 &  ACS    & 1.0, 1.0  & 904, 1561    & 1375, 1306 &  1.0, 1.0     &  .99, .99  &  0.006  &  \\
   &  DC             &  SAS    & 1.0, 1.0  & 4800, 4800   & 1072, 393  &  1.0, 1.0     &  .99, .99  &  0.459  &12586\\
   &                 &  ACS    & .88, .96  & 5, 6         & 61, 108    &  1.0, 1.0     &  .20, .33  &  0.459  & \\
\hline
120&  Pearson        &  SAS    & 1.0, 1.0  & 4800, 4800   & 0, 0       &  1.0, 1.0     &  .99, .99  &  0.006 &0.172  \\
   &                 &  ACS    & 1.0, 1.0  & 904, 1561    & 1375, 1306 &  1.0, 1.0     &  .99, .99  &  0.006 & \\
   &  DC             &  SAS    & 1.0, 1.0  & 4800, 4800   & 0, 0       &  1.0, 1.0     &  .99, .99  &  0.190 &12586 \\
   &                 &  ACS    & .83, .95  & 6, 13        & 111, 189   &  1.0, 1.0     &  .33, .68  &  0.190 & \\
\hline
\multicolumn{10}{c}{Model (f), $N=10000$, $p=10000$, $\|\mathcal{M}\|_{0}=4$} \\
\hline
100&  Pearson        &  SAS    & 1.0, 1.0  & 10000, 10000 & 0, 0       &  1.0, 1.0     &  .99, .99  &  0.016 &0.729  \\
   &                 &  ACS    & 1.0, 1.0  & 8600, 8950   & 3365, 2946 &  1.0, 1.0     &  .99, .99  &  0.016 & \\
   &  DC             &  SAS    & 1.0, 1.0  & 10000, 10000 & 1809, 0    &  1.0, 1.0     &  .99, .99  &  1.923 &213998 \\
   &                 &  ACS    & .98, 1.0  & 8, 8         &  1, 2      &  1.0, 1.0     &  .50, .50  &  1.923 & \\
\hline
250&  Pearson        &  SAS    & 1.0, 1.0  & 10000, 10000 & 0, 0       &  1.0, 1.0     &  .99, .99  &  0.017 &0.729 \\
   &                 &  ACS    & 1.0, 1.0  & 8600, 8950   & 3365, 2946 &  1.0, 1.0     &  .99, .99  &  0.017 & \\
   &  DC             &  SAS    & 1.0, 1.0  & 10000, 10000 & 0, 0       &  1.0, 1.0     &  .99, .99  &  0.441 &213998  \\
   &                 &  ACS    & .85, 1.0  & 8, 10        & 9, 50      &  1.0, 1.0     &  .50, .60  &  0.441 & \\
\hline
\hline
\end{tabular*}
\end{table}

With the ``oracle" choice of $\gamma$, we see that both SAS and ACS perform well in terms of keeping relevant features; this is indicated by their high SSRs in most cases. Regarding the screening precision, SAS seems to be inferior, as it tends to keep too many irrelevant features after screening. This phenomenon is particularly severe for non-linear correlation measures SIRS and DC under the large-$m$-small-$\gamma$ setup. As an extreme case, when DC is used in Model (e), SAS suggests keeping all the 4800 features; this completely fails in the mission of screening. The over-selection of SAS here is a direct result from the inaccuracy of the corresponding simple averaging estimators $\bar{\omega}_{j}$s. When Pearson and Kendall $\tau$ correlations are used, this issue is less severe, as the corresponding $\bar{\omega}_{j}$s are less biased due to their nature. In addition, we observe a high variability for SAS-based screening in most cases; this makes it less trustable in practice. The proposed ACS, in comparison, is built upon the stable $\widetilde{\omega}_j$s, and thus achieves a reasonably high screening precision in most setups. For all the four correlation choices, it is able to screen most irrelevant features out, while keep relevant ones with a high probability. Such a performance is very promising.

We observe that, when SIRS is used in Model (a) with $m=30$, neither ACS nor SAS works very well, if SSR and screening precision are considered jointly. This might be due to the relatively low sensitivity of SIRS in detecting linear correlations when $n$ is small. Apparently, using multiple data partition strategy helps a lot in this case, as indicated by the high SSR and low Std(MS) of the corresponding rACS.

Benefited from its distributed framework, the proposed ACS enables parallel computing and enjoys a great numerical advantage over the classic screening procedures (i.e. $m=1$ case). As shown in Tables \ref{Table2}-\ref{Table4}, the computational cost of ACS can be even less than 1\% of the traditional cost with a large $m$ setup, while it still maintains relatively high screening accuracy. This merit together with its broad compatibility makes ACS an attractive approach for screening with large-$N$-large-$p$ data.

\subsection{A real data analysis}\label{sec3-1}

We apply the proposed ACS to a real dataset\footnote{Available at http://archive.ics.uci.edu/ml/datasets/Superconductivty+Data}, which contains 81 covariates extracted from 21,263 superconductors along with the associated critical temperature (response). Readers may refer to \cite{hamidieh2018data} for a detailed description of this dataset. It is of interest to predict the unknown response given a set of new values of the covariates. It is likely that the covariates are linked to the response with a non-linear relationship. To avoid potential model mis-specification, we build a non-parametric kernel ridge regression of the critical temperature on the full data input.

Specifically, let $(y_i, \boldsymbol{x}_i)$ denote the critical temperature and the corresponding 81-dimensional covariate vector for the $i$th superconductor.
We seek for a predictive function $\hat{f}$ by minimizing
\[
\hat{f}= \arg\min \limits_{f} \left\{\frac{1}{N}\sum_{i=1}^N (y_i- f(\boldsymbol{x}_i))^2+\lambda \|f\|_{K}^2\right\},
\]
where $f$ has the form
\[
{f}(\boldsymbol{x}) = \sum_{j=1}^{N}\beta_j K(\boldsymbol{x}, \boldsymbol{x}_j),
\]
$\|f\|^{2}_{K}= \sum_{i,j=1}^{N}\beta_i\beta_j K(\boldsymbol{x}_i,\boldsymbol{x}_{j})$ is the norm of $f$ induced by a user-specified kernel function $K$, and
$\lambda>0$ is a tuning parameter. In this numerical study, we choose the gaussian kernel $K(\boldsymbol{x}_i, \boldsymbol{x}_j)=\exp(-\|\boldsymbol{x}_i-\boldsymbol{x}_j\|_2^2/\phi^2)$ with $\phi=10$.  We remove data entries with missing values in $\boldsymbol{x}_i$ and get 20,877 available data entries, from which we randomly select $20,000$ entries as a training set and treat the remaining 877 ones as a testing set.
This leads to a working kernel matrix $\textbf{K}=\{K(\boldsymbol{x}_i,\boldsymbol{x}_{j}), \ i,j = 1, \ldots, N\}$ with $N=20,000$ observations and $p= N =20,000$ kernel atoms $K(\cdot, \boldsymbol{x}_{j})$ that are evaluated at each observation $\boldsymbol{x}_{i}$ in the training set.

\begin{figure}[t]
\setlength{\parindent}{-2.5em} \centering
\scalebox{0.65}{\includegraphics{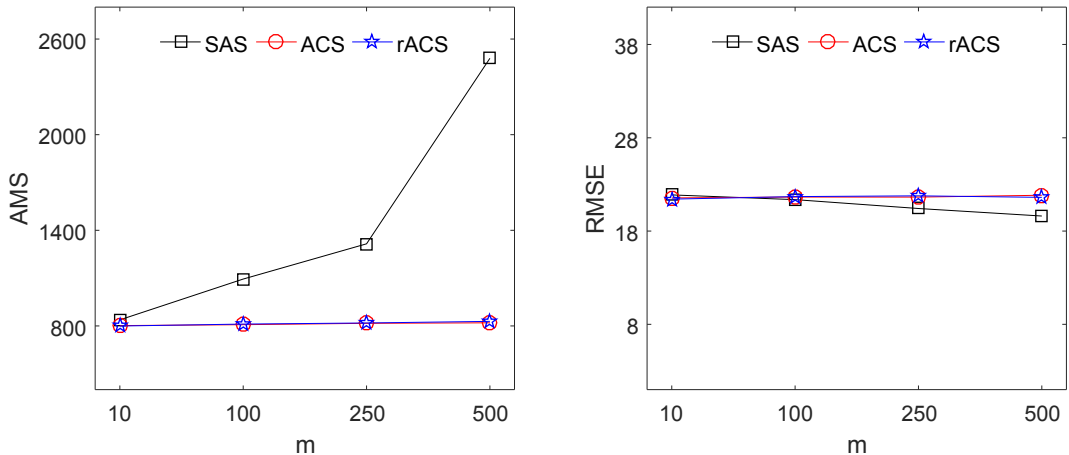}}
\caption{ Analysis of superconductor data. }\label{fig2}
\end{figure}

Apparently, $\textbf{K}$ is likely to contain a large number of redundant kernel atoms (features) that are irrelevant for prediction. We apply the proposed ACS with SIRS for a model-free feature screening. Specifically, we randomly partition the training set into $m=10, 100, 200, 500$ segments and run ACS as well as rACS with $R=3$ respectively for each case. The screening threshold $\gamma$ is set by the $800$th largest $\widetilde{\omega}_j$ in rACS with $m=10$. We evaluate the screening results in terms of the averaged model size (AMS) as well as the averaged prediction error (RMSE) of the corresponding $\hat{f}$ based on the testing set, where the $\lambda$ in $\hat{f}$ is determined by 10-fold cross validation. We summarize the results in Figure \ref{fig2} with 100 repetitions, where the performance of SAS is also reported.

In this example, it seems that all screening methods under consideration lead to a similar predictive accuracy. When $m$ is small, SAS and ACS tend to keep the same amount of relevant features. As $m$ increases, SAS becomes more liberal by retaining more features after screening, while ACS remains restrictive. When $m=500$, SAS suggests 2479 ``relevant" features, which is about 3 times the number of features suggested by ACS. Yet, as indicated by their RMSEs, including a large number of features in $\hat{f}$ does not help to significantly improve the predictive power. This implies that a large portion of the SAS-suggested features are actually redundant. In comparison,  ACS is accurate and stable among all $m$ setups and thus leads to a more reliable screening result in general.

\section{Concluding Remarks}\label{sec5}

Technological innovations have made a profound impact on knowledge discovery. Extracting useful features from massive amount of high dimensional data is essential in many modern scientific areas. In this paper, we proposed a distributed framework (ACS) for feature screening with large-$N$-large-$p$ datasets. In the spirit of ``divide-and-conquer", ACS enables distributed storage and paralleling computing, and thus enjoys a great numerical advantage over the classic screening methods. The key of success for ACS is that we express a correlation measure as a function of several component parameters and conduct distributive unbiased estimation for each of them. With the unbiased component estimates combined together, we then obtained an aggregated correlation estimate $\widetilde{\omega}_{j}$, which is accurate and insensitive to the number data segments used in the analysis. This further leads to a computationally efficient and performance reliable screening procedure. Under mild conditions, we showed that $\widetilde{\omega}_{j}$ is as efficient as the classic centralized estimators, while it drastically reduces the computational cost. The corresponding screening procedure is compatible with a broad range of correlation measures and enjoys the desirable sure screening property.

It should be noted that our current discussion is based on the i.i.d assumption of $(Y_{i}, \mathbf{X}_{i})$, which can be impractical when data segments are naturally stored at different locations. In such a scenario, it is likely that data segments are of different sizes and qualities. To make the proposed ACS more adaptive, one may replace $\bar{U}_{j,h}$ in (\ref{agg u}) by a weighted average, where the weight is proportional to the inverse-variance of the local component estimator $U^l_{j,h}$. We leave this interesting work for future research.

\section{Acknowledgement}\label{sec6}
Xu's research was supported by NSERC grant RGPIN-2016-05024 and NSFC grant 116900 14.
Runze Li's research was supported by NSF grant DMS 1820702 and NIDA, NIH grant P50 DA039838.
Xia's research was supported by NSFC grant 11771353.
The content is solely the responsibility of the authors and does not necessarily represent the official views of the aforementioned funding agencies.

\section*{Appendix}
\noindent
{\bf Proof of Proposition \ref{prop1}.}
Let $\hat{\theta}_{j,h}$ be a basis unbiased estimator of $\theta_{j,h}$ with degree $k_{h}$. By Markov's inequality,  we have     
\begin{eqnarray}\label{Prop-1}
P(\bar{U}_{j,h}-\theta_{j,h} \geq \varepsilon) &=& P( \exp\{ \nu(\bar{U}_{j,h}-\theta_{j,h}) \} \geq \exp\{ \nu \varepsilon\} )   \nonumber  \\
                           &\leq& \exp\{-\nu\varepsilon\}\exp\{-\nu\theta_{j,h}\}E[\exp\{\nu\bar{U}_{j,h}\}],
\end{eqnarray}
for any $\varepsilon>0$ and $0< \nu \leq\kappa_0mr_h$ with $r_h=\lfloor n/k_h\rfloor$.

Let $\mathcal{S}_l=\{l_1,...,l_n\}$ denote the index set of $\{Y, \textbf{X}\}$ copies based on $\mathcal{D}_{l}$, on which we can construct $r_h$ independent $\hat{\theta}_{j,h}$s. We define an averaged estimator based on those $\hat{\theta}_{j,h}$s by
$$
V_{j,h}(Z_{l_1j},...,Z_{l_nj})=\frac{1}{r_h}\sum_{u=1}^{r_h}\hat{\theta}_{j,h}(Z_{l_{(u-1)k_h+1}j},...,Z_{l_{uk_h}j}).
$$
Then, the local U-statistic in (\ref{local u}) can be expressed by
$${U}^{l}_{j,h}=\frac{1}{n!}\sum_{\{i_1,...,i_n\}\in \Omega}V_{j,h}(Z_{l_{i_1}j},...,Z_{l_{i_n}j}),$$
where $\Omega=\{1,...,n\}$ and the summation is over all $\{Z_{l_{i_1}j},...,Z_{l_{i_n}j}\}$ permutations from $\mathcal{D}_l$.
Consequently,
\begin{eqnarray*}
\bar{U}_{j,h}=\frac{1}{m}\sum_{l=1}^{m}U^{l}_{j,h}=\frac{1}{n!}\sum_{\{i_1,...,i_n\}\in \Omega}
\frac{1}{m}\sum_{l=1}^{m}V_{j,h}(Z_{l_{i_1}j},...,Z_{l_{i_n}j}).
\end{eqnarray*}
Since exponential function is convex, Jensen's inequality implies that
\begin{eqnarray}\label{Prop-2}
E[\exp\{\nu\bar{U}_{j,h}\}]&=&E\left[\exp\left\{\frac{\nu}{n!}\sum_{\{i_1,...,i_n\}\in \Omega}\left(\frac{1}{m}\sum_{l=1}^{m}V_{j,h}(Z_{l_{i_1}j},...,Z_{l_{i_n}j})\right)\right\}\right] \nonumber\\
&\leq& \frac{1}{n!}\sum_{\{i_1,...,i_n\}\in \Omega}E\left[\exp\left\{\frac{\nu}{m}\sum_{l=1}^{m}V_{j,h}(Z_{l_{i_1}j},...,Z_{l_{i_n}j})\right\}\right] \nonumber\\
&=& \psi_{j,h}^{mr_h}\left(\kappa\right),
\end{eqnarray}
where $\kappa=\nu/(mr_h)$ and $\psi_{j,h}(\kappa)=E[\exp\{\kappa\hat{\theta}_{j,h}\}]$.

Combining (\ref{Prop-1}) and (\ref{Prop-2}), we have
\begin{eqnarray}\label{Prop-3}
P(\bar{U}_{j,h}-\theta_{j,h} \geq \varepsilon) \leq [\exp\{-\kappa\varepsilon\}\exp\{-\kappa\theta_{j,h}\}\psi_{j,h}(\kappa)]^{mr_h}.
\end{eqnarray}

Let $V$ be a generic variable. By Taylor expansion, we have $\exp\{\kappa V\}=1+\kappa V+\kappa^2V'/2$, where $0 < V' < V^2\exp\{\kappa_1V\}$ for some $\kappa_1 \in (0, \kappa)$. Thus, factor $\exp\{-\kappa\theta_{j,h}\}\psi_{j,h}(\kappa)$ in (\ref{Prop-3}) can be bounded by
\begin{eqnarray}
\exp\{-\kappa\theta_{j,h}\}\psi_{j,h}(\kappa)&=&E [ \exp\{\kappa(\hat{\theta}_{j,h}-\theta_{j,h})\} ]\nonumber\\
&=& E\left[1+\kappa(\hat{\theta}_{j,h}-\theta_{j,h})+\kappa^2\exp\{\kappa_1(\hat{\theta}_{j,h}-\theta_{j,h})\}(\hat{\theta}_{j,h}-\theta_{j,h})^2/2\right] \nonumber\\
&=& 1+\kappa^2E\left[(\hat{\theta}_{j,h}-\theta_{j,h})^2\exp\{\kappa_1(\hat{\theta}_{j,h}-\theta_{j,h})\}\right]/2 \nonumber\\
&\leq& 1+\kappa^2[E\hat{\theta}_{j,h}^4\cdot E\exp\{2\kappa_1(\hat{\theta}_{j,h}-\theta_{j,h})\}]^{1/2}/2,\label{Prop-4}
\end{eqnarray}
where (\ref{Prop-4}) is implied by H$\ddot{\textnormal{o}}$lder's inequality.

By Condition C1, we know (\ref{Prop-4}) can be bounded by $1+D_1\kappa^2$ with some $D_1>0$. Also, when $\kappa \varepsilon < 1$,  we have $\exp(-\kappa\varepsilon)\leq 1-\varepsilon \kappa+D_2\varepsilon^{2}\kappa^2$ with some $D_2>0$. Thus, we have the base term in (\ref{Prop-3}) bounded by
\begin{eqnarray*}
\exp\{-\kappa\varepsilon\}\exp\{-\kappa\theta_{j,h}\}\psi_{j,h}(\kappa) &\leq& (1+D_1\kappa^2)(1-\varepsilon \kappa+D_2\varepsilon^{2}\kappa^2) \\
&=& 1 - \varepsilon \kappa + D_2\kappa^2\varepsilon^2+D_1\kappa^2-D_1\kappa^3\varepsilon+D_1D_2\kappa^4\varepsilon^2\\
&\leq& 1 - \varepsilon \kappa + D_2\kappa^2\varepsilon^2+D_1\kappa^2+D_1D_2\kappa^4\varepsilon^2 \\
&=& 1 - \varepsilon \kappa + {E}_1,
\end{eqnarray*}
where ${E}_1=D_2\kappa^2\varepsilon^2+D_1\kappa^2+D_1D_2\kappa^4\varepsilon^2$.
By setting $\kappa=c_0\varepsilon$, we have
\begin{eqnarray}\label{prop-c0}
\frac{{E}_1}{\kappa\varepsilon}&=&D_2c_0\varepsilon^2+D_1c_0+D_1D_2c_0^3\varepsilon^4 \nonumber\\
&\leq& D_2c_0\delta_0^2+D_1c_0+D_1D_2c_0^3\delta_0^4 .
\end{eqnarray}
Note that, when $c_0>0$ is small enough, we have $\kappa \in (0, \kappa_0)$, $\kappa \varepsilon < 1$, and (\ref{prop-c0}) is bounded by $1/2$.
Thus,  the base term in (\ref{Prop-3}) is further bounded  by      
\begin{eqnarray}\label{Prop-6}
\exp\{-\kappa\varepsilon\}\exp\{-\kappa\theta_{j,h}\}\psi_{j,h}(\kappa)\leq 1-\varepsilon \kappa/2.
\end{eqnarray}
Combining (\ref{Prop-3}) and (\ref{Prop-6}),  we have
$$
P(\bar{U}_{j,h}-\theta_{j,h}\geq \varepsilon)\leq (1-c_0\varepsilon^2/2)^{mr_h}.
$$

Similarly, we can show that $P(\bar{U}_{j,h}-\theta_{j,h}\leq -\varepsilon)\leq(1-c_0\varepsilon^2/2)^{mr_h}$.
Therefore, we obtain
\begin{eqnarray*}
P(|\bar{U}_{j,h}-\theta_{j,h}|\geq \varepsilon)\leq  2(1-c_0\varepsilon^2/2)^{m\lfloor n/k_h\rfloor}
\end{eqnarray*}
under the conditions specified in the proposition. The proof is complete. \hfill$\Box$

\vspace{0.4cm}
The proof of Theorem \ref{Thm1} is built upon the following technical lemma.

\vspace{0.2cm}
\noindent
\begin{lemma}\label{lemma1}
Suppose that $\theta_h, h=1,...,s$ are bounded, that is, there exists a positive constant $a>0$ such that $|\theta_h|<a$. Let $\widetilde{\theta}_h$ be an estimator of $\theta_h$. Suppose for any $\varepsilon\in(0,c]$, there exists a constant $c_1>0$ such that, for any  $h\in\{1,...,s\}$,
\begin{eqnarray}\label{Lem-1}
 P(|\widetilde{\theta}_h-\theta_h|\geq\varepsilon)\leq c_1(1-\varepsilon^2/c_1)^{m\lfloor n/k\rfloor},
\end{eqnarray}
where $k$ is a positive integer. Then, there exists a positive constant $c'$ such that
\begin{eqnarray}
P\left(\left||\widetilde{\theta}_{h}|-|\theta_{h}|\right|\geq\varepsilon\right)&\leq& c'(1-\varepsilon^2/c')^{m\lfloor n/k \rfloor}, \label{Lem-2}\\
P(|(\widetilde{\theta}_{h_1}+\widetilde{\theta}_{h_2})-(\theta_{h_1}+\theta_{h_2})|\geq\varepsilon)&\leq&
c'(1-\varepsilon^2/c')^{m\lfloor n/k\rfloor},\label{Lem-3}\\
P(|(\widetilde{\theta}_{h_1}-\widetilde{\theta}_{h_2})-(\theta_{h_1}-\theta_{h_2})|\geq\varepsilon)
&\leq& c'(1-\varepsilon^2/c')^{m\lfloor n/k \rfloor}, \label{Lem-4}\\
P(|\widetilde{\theta}_{h_1}\widetilde{\theta}_{h_2}-\theta_{h_1}\theta_{h_2}|\geq\varepsilon)
&\leq& c'(1-\varepsilon^2/c')^{m\lfloor n/k \rfloor},\label{Lem-5}\\
P(|\widetilde{\theta}_{h}^2-\theta_{h}^2|\geq\varepsilon)
&\leq& c'(1-\varepsilon^2/c')^{m\lfloor n/k \rfloor}.\label{Lem-6}
\end{eqnarray}
Moreover, suppose there exists a constant $b>0$ such that $|\theta_{h_2}|>b$. Then, we have
\begin{eqnarray}
P(|\widetilde{\theta}_{h_1}/\widetilde{\theta}_{h_2}-\theta_{h_1}/\theta_{h_2}|\geq\varepsilon)&\leq& c'(1-\varepsilon^2/c')^{m\lfloor n/k \rfloor}. \label{Lem-7}
\end{eqnarray}
If we further assume $\theta_{h}> 0$, then
\begin{eqnarray}
P\left(\left|\sqrt{\widetilde{\theta}_{h}}-\sqrt{\theta_{h}}\right|\geq\varepsilon\right)&\leq& c'(1-\varepsilon^2/c')^{m\lfloor n/k \rfloor}. \label{Lem-8}
\end{eqnarray}
\end{lemma}

\vspace{0.5cm}
\noindent
{\bf Proof of Lemma \ref{lemma1}.} We prove the lemma by justifying (\ref{Lem-2})-(\ref{Lem-8}) sequentially.

The proof of (\ref{Lem-2}) is straightforward. By (\ref{Lem-1}), for any $\varepsilon\in(0,c]$, we have
\begin{eqnarray*}
P\left(\left||\widetilde{\theta}_h|-|\theta_h|\right|\geq\varepsilon \right)&\leq&P\left(\left|\widetilde{\theta}_h-\theta_h\right|\geq\varepsilon \right) \nonumber\\
&\leq& c_1(1-\varepsilon^2/c_1)^{m\lfloor n/k\rfloor}.
\end{eqnarray*}

\vspace{0.5cm}
We now work on (\ref{Lem-3}). For any $\varepsilon\in(0,c]$, we have
\begin{eqnarray*}
&&P(|(\widetilde{\theta}_{h_1}+\widetilde{\theta}_{h_2})-(\theta_{h_1}+\theta_{h_2})|\geq\varepsilon)\nonumber\\
&\leq&P(|\widetilde{\theta}_{h_1}-\theta_{h_1}|\geq\varepsilon/2)+P(|\widetilde{\theta}_{h_2}-\theta_{h_2}|\geq\varepsilon/2)\nonumber\\
&\leq&2c_1(1-\varepsilon^2/(4c_1))^{m\lfloor n/k\rfloor} \leq c_2(1-\varepsilon^2/c_2)^{m\lfloor n/k \rfloor},
\end{eqnarray*}
where $c_2=4c_1$. Similarly, we can also show (\ref{Lem-4}).

\vspace{0.5cm}
To show (\ref{Lem-5}), we first prove that $\widetilde{\theta}_h$s are bounded in probability. Specifically, since $|\theta_h|\leq a$, we have, for any $\varepsilon\in(0,c]$,
\begin{eqnarray}
P\left(|\widetilde{\theta}_h|\geq a+\varepsilon \right)&\leq& P\left(|\widetilde{\theta}_h-\theta_h|+|\theta_h|\geq a+\varepsilon \right)\nonumber\\
&\leq& P\left(|\widetilde{\theta}_h-\theta_h|\geq \varepsilon \right)\nonumber\\
&\leq& c_1(1-\varepsilon^2/c_1)^{m\lfloor n/k\rfloor}. \label{Lem-9}
\end{eqnarray}
Therefore,
\begin{eqnarray}
&&P(|\widetilde{\theta}_{h_1}\widetilde{\theta}_{h_2}-{\theta}_{h_1}{\theta}_{h_2}|\geq\varepsilon)\nonumber\\
&\leq& P(|\widetilde{\theta}_{h_1}\widetilde{\theta}_{h_2}-\widetilde{\theta}_{h_1}{\theta}_{h_2}+\widetilde{\theta}_{h_1}{\theta}_{h_2}-{\theta}_{h_1}{\theta}_{h_2}|\geq\varepsilon)\nonumber\\
&\leq& P(|\widetilde{\theta}_{h_1}|\cdot|\widetilde{\theta}_{h_2}-{\theta}_{h_2}|+|{\theta}_{h_2}|\cdot|\widetilde{\theta}_{h_1}-{\theta}_{h_1}|\geq\varepsilon)\nonumber\\
&\leq& P(|\widetilde{\theta}_{h_1}|\cdot|\widetilde{\theta}_{h_2}-{\theta}_{h_2}|\geq\varepsilon/2)+P(|{\theta}_{h_2}|\cdot|\widetilde{\theta}_{h_1}-{\theta}_{h_1}|\geq\varepsilon/2).\label{Lem-10}
\end{eqnarray}
By (\ref{Lem-1}) and (\ref{Lem-9}), the first term of (\ref{Lem-10}) can be bounded by
\begin{eqnarray*}
&&P(|\widetilde{\theta}_{h_1}|\cdot|\widetilde{\theta}_{h_2}-{\theta}_{h_2}|\geq\varepsilon/2)\nonumber\\
&=&P(|\widetilde{\theta}_{h_1}|\cdot|\widetilde{\theta}_{h_2}-{\theta}_{h_2}|\geq\varepsilon/2,|\widetilde{\theta}_{h_1}|\geq a+\varepsilon)\nonumber\\
&&+P(|\widetilde{\theta}_{h_1}|\cdot|\widetilde{\theta}_{h_2}-{\theta}_{h_2}|\geq\varepsilon/2,|\widetilde{\theta}_{h_1}|< a+\varepsilon)\nonumber\\
&\leq&P(|\widetilde{\theta}_{h_1}|\geq a+\varepsilon)+P((a+\varepsilon)\cdot|\widetilde{\theta}_{h_2}-{\theta}_{h_2}|\geq\varepsilon/2)\nonumber\\
&\leq& c_1(1-\varepsilon^2/c_1)^{m\lfloor n/k\rfloor}+c_1(1-\varepsilon^2/c_3)^{m\lfloor n/k \rfloor},
\end{eqnarray*}
where $c_3=\max\{4(a+c)^2c_1, c_1\}$. The second term of (\ref{Lem-10}) can be bounded by
\begin{eqnarray*}
P(|{\theta}_{h_2}|\cdot|\widetilde{\theta}_{h_1}-{\theta}_{h_1}|\geq\varepsilon/2)&\leq& P(|\widetilde{\theta}_{h_1}-{\theta}_{h_1}|\geq\varepsilon/(2a))\nonumber\\
&\leq&  c_1(1-\varepsilon^2/c_4)^{m\lfloor n/k \rfloor}
\end{eqnarray*}
with $c_4=\max\{4a^2c_1,c_1\}$. Then, by setting $c_5=\max\{3c_1,c_3\}$, we have
\begin{eqnarray*}
P(|\widetilde{\theta}_{h_1}\widetilde{\theta}_{h_2}-{\theta}_{h_1}{\theta}_{h_2}|\geq\varepsilon)\leq
3c_1(1-\varepsilon^2/c_3)^{m\lfloor n/k \rfloor}\leq
c_5(1-\varepsilon^2/c_5)^{m\lfloor n/k \rfloor},
\end{eqnarray*}
which proves (\ref{Lem-5}).
By setting $\widetilde{\theta}_{h_2} = \widetilde{\theta}_{h_1}=\widetilde{\theta}_{h}$ in (\ref{Lem-5}), we immediately have result (\ref{Lem-6}).

\vspace{0.5cm}
To prove (\ref{Lem-7}), let us first show that $\widetilde{\theta}_{h_2}$ is bounded away from 0 in probability. Since $|{\theta}_{h_2}|>b>0$, there exists a constant $\delta_1\in (0,c)$ such that
for some $b' = b-\delta_1>0$,
\begin{eqnarray*}
P(|\widetilde{\theta}_{h_2}|\leq b')&\leq& P(|\theta_{h_2}|-|\widetilde{\theta}_{h_2}-{\theta}_{h_2}|\leq b-\delta_1)\nonumber\\
&\leq& P(|\widetilde{\theta}_{h_2}-{\theta}_{h_2}|\geq \delta_1)\nonumber\\
&\leq& c_1(1-\delta_1^2/c_1)^{m\lfloor n/k \rfloor}.
\end{eqnarray*}
Let $c_6=c_1c^2/\delta_1^2$. Then, for $\varepsilon\in(0,c)$, we have
\begin{eqnarray}
P(|\widetilde{\theta}_{h_2}|\leq b')\leq c_1(1-\varepsilon^2/{c_6})^{m\lfloor n/k \rfloor}. \label{Lem-11}
\end{eqnarray}
 Based on (\ref{Lem-11}), we have
\begin{eqnarray}
&&P(|\widetilde{\theta}_{h_1}/\widetilde{\theta}_{h_2}-\theta_{h_1}/\theta_{h_2}|\geq\varepsilon)\nonumber\\
&=&P(|\widetilde{\theta}_{h_1}/\widetilde{\theta}_{h_2}-\theta_{h_1}/\theta_{h_2}|\geq\varepsilon, |\widetilde{\theta}_{h_2}|\leq
b')+P(|\widetilde{\theta}_{h_1}/\widetilde{\theta}_{h_2}-\theta_{h_1}/\theta_{h_2}|\geq\varepsilon, |\widetilde{\theta}_{h_2}|> b')\nonumber\\
&\leq&P(|\widetilde{\theta}_{h_2}|\leq b')+P(|\widetilde{\theta}_{h_1}/\widetilde{\theta}_{h_2}-\theta_{h_1}/\theta_{h_2}|\geq\varepsilon, |\widetilde{\theta}_{h_2}|> b')\nonumber\\
&\leq&c_1(1-\varepsilon^2/{c_6})^{m\lfloor n/k \rfloor}+P(|\widetilde{\theta}_{h_1}/\widetilde{\theta}_{h_2}-\theta_{h_1}/\theta_{h_2}|\geq\varepsilon, |\widetilde{\theta}_{h_2}|> b').\label{Lem-12}
\end{eqnarray}
In (\ref{Lem-12}), the second term can be bounded by
\begin{eqnarray}
&&P(|\widetilde{\theta}_{h_1}/\widetilde{\theta}_{h_2}-\theta_{h_1}/\theta_{h_2}|\geq\varepsilon, |\widetilde{\theta}_{h_2}|> b')\nonumber\\
&\leq&P(|\widetilde{\theta}_{h_1}/\widetilde{\theta}_{h_2}-\theta_{h_1}/\widetilde{\theta}_{h_2}|+|{\theta}_{h_1}/\widetilde{\theta}_{h_2}-\theta_{h_1}/\theta_{h_2}|\geq\varepsilon, |\widetilde{\theta}_{h_2}|>b')\nonumber\\
&\leq&P\left(\frac{1}{b'}|\widetilde{\theta}_{h_1}-\theta_{h_1}|\geq\varepsilon/2\right)+P\left(\frac{|{\theta}_{h_1}|}{|\widetilde{\theta}_{h_2}|\cdot|{\theta}_{h_2}|}|\widetilde{\theta}_{h_2}-\theta_{h_2}|\geq\varepsilon/2\right)\nonumber\\
&\leq&c_1(1-\varepsilon^2/c_7)^{m\lfloor n/k \rfloor}+c_1(1-\varepsilon^2/c_8)^{m\lfloor n/k \rfloor},\label{Lem-13}
\end{eqnarray}
where $c_7=\max\{4c_1/(b')^2,c_1\}$ and $c_8=\max\{4a^2c_1/(b'b)^2,c_1\}$. Let $c_9=\max\{3c_1, c_6, c_7, c_8\}$, then we have
\begin{eqnarray*}
&&P(|\widetilde{\theta}_{h_1}/\widetilde{\theta}_{h_2}-\theta_{h_1}/\theta_{h_2}|\geq\varepsilon) \leq  c_9(1-\varepsilon^2/c_9)^{m\lfloor n/k \rfloor}.
\end{eqnarray*}

\vspace{0.5cm}
Lastly, let us work on (\ref{Lem-8}). Since $\theta_{h}>0$, there exists a $\tilde{b}>0$ such that $\theta_{h}>\tilde{b}$. Similar to (\ref{Lem-11})-(\ref{Lem-13}), there exist two positive constants $\tilde{b}'$ and $c_{10}$ such that
\begin{eqnarray*}
&&P\left(\left|\sqrt{\widetilde{\theta}_{h}}-\sqrt{\theta_{h}}\right|\geq\varepsilon\right)\nonumber\\
&\leq& P(|\widetilde{\theta}_{h}|\leq \tilde{b}')+P\left(\frac{\widetilde{\theta}_{h}-{\theta}_{h}}{\sqrt{\widetilde{\theta}_{h}}+\sqrt{{\theta}_{h}}}\geq \varepsilon, |\widetilde{\theta}_{h}|>\tilde{b}'\right)\nonumber\\
&\leq& c_1(1-\varepsilon^2/{c_{10}})^{m\lfloor n/k \rfloor}+P\left(|\widetilde{\theta}_{h}-{\theta}_{h}|\geq (\sqrt{\tilde{b}'}+\sqrt{\tilde{b}})\varepsilon \right)\nonumber\\
&\leq& c_1(1-\varepsilon^2/{c_{10}})^{m\lfloor n/k \rfloor} + c_1(1-\varepsilon^2/c_{11})^{m\lfloor n/k \rfloor},
\end{eqnarray*}
where $c_{11}= \max\{{c_1}/(\sqrt{\tilde{b}'}+\sqrt{\tilde{b}})^2,c_1\}$. By setting $c_{12}=\max\{2c_1, c_{10}, c_{11}\}$, we obtain that
\begin{eqnarray*}
P\left(\left|\sqrt{\widetilde{\theta}_{h}}-\sqrt{\theta_{h}}\right|\geq\varepsilon\right)\leq c_{12}(1-\varepsilon^2/c_{12})^{m\lfloor n/k \rfloor}.
\end{eqnarray*}
Result (\ref{Lem-8}) is therefore proved.

\vspace{0.2cm}
Combining the results in (\ref{Lem-2})-(\ref{Lem-8}),  we prove Lemma \ref{lemma1} by setting $c'=\max\{c_1,c_2,c_5,$ $c_9,c_{12}\}$.   \hfill$\Box$

\vspace{0.3cm}

With Proposition \ref{prop1} and Lemma \ref{lemma1}, we prove Theorem \ref{Thm1} as follows.

\vspace{0.2cm}
\noindent
{\bf Proof of Theorem \ref{Thm1}.} By Proposition \ref{prop1}, for any $\varepsilon\in(0,\delta_0]$, there exists a
$c_0>0$ such that
\begin{eqnarray*}
 P(|\bar{U}_{j,h}-\theta_{j,h}|\geq\varepsilon)\leq 2(1-c_0\varepsilon^2)^{m\lfloor n/k_h\rfloor}\leq 2(1-c_0\varepsilon^2)^{m\lfloor n/k \rfloor},
\end{eqnarray*}
where $k= \max\{k_{h}, h= 1, \ldots, s\} \leq n$.
Since $\delta_0$ can be arbitrarily large, the inequality holds with $\varepsilon=cN^{-\tau}\in(0,c]$ for some $0 < \tau < 1/2$.
Thus, we have
\begin{eqnarray} \label{Thm1-1}
P(|\bar{U}_{j,h}-\theta_{j,h}|
\geq cN^{-\tau})
&\leq &
2(1-c_{13}N^{-2\tau}/2)^{m\lfloor n/k \rfloor}\nonumber\\
&\leq &c_{14}(1-N^{-2\tau}/c_{14})^{m\lfloor n/k \rfloor}, \ \ h=1,...,s,
\end{eqnarray}
where $c_{14}=\max\{2, 2/c_{13}\}$.
This implies that the results of Lemma 1 are applicable by setting $\widetilde{\theta}_{h} = \bar{U}_{j,h}$.

By Condition 2, we require that $\widetilde{\omega}_j = g(\bar{U}_{j,1}, \ldots, \bar{U}_{j,s})$ is constructed by a finite number of simple numerical operations, which serve as basic building blocks of $g(\cdot)$.  For each building block, Lemma \ref{lemma1} can be used immediately to establish the convergence bound for the corresponding $\widetilde{\omega}_j$.  With finite combination of those building blocks, (\ref{Thm1-1}) further implies that
\begin{eqnarray*}
P(|\widetilde{\omega}_j-\omega_j|\geq cN^{-\tau})\leq \eta(1-N^{-2\tau}/\eta)^{m\lfloor n/k \rfloor}
\end{eqnarray*}
for some generic positive constant $\eta$.

Consequently, we have
\begin{eqnarray*}
P\left(\max \limits_{1 \leq j \leq p} |\widetilde{\omega}_j-\omega_j|\geq cN^{-\tau} \right)&\leq& \sum_{j=1}^p  P\left(|\widetilde{\omega}_j-\omega_j|\geq cN^{-\tau} \right)\nonumber\\
&\leq& \eta p (1-N^{-2\tau}/\eta)^{m\lfloor n/k \rfloor}.
\end{eqnarray*}
The theorem is proved. \hfill$\Box$

\vspace{0.5cm}
\noindent
{\bf Proof of Theorem \ref{Thm2}.}
Note that $\gamma =cN^{-\tau}$. If $\mathcal {M}\nsubseteq \widetilde{\mathcal {M}}$, there must exist some $j\in \mathcal {M}$ such that $\widetilde{\omega}_j<cN^{-\tau}$. Also, by Condition C3, we assume $ \min \limits_{j \in \mathcal {M}} \omega_j \geq 2cN^{-\tau}$. Thus, $\mathcal{M}\nsubseteq\widetilde{\mathcal {M}}$ implies $|\widetilde{\omega}_j-\omega_j|>cN^{-\tau}$ for some $j\in \mathcal {M}$. Therefore, we have
\begin{eqnarray*}
P\{ \mathcal{M}\subseteq\widetilde{\mathcal {M}} \}&\geq& P(\max \limits_{j \in \mathcal {M}} \left|\widetilde{\omega}_j-\omega_j\right| \leq cN^{-\tau}) \nonumber\\
&\geq&1-P(\max \limits_{j \in \mathcal {M}} |\widetilde{\omega}_j-\omega_j|> cN^{-\tau}) \nonumber\\
&\geq& 1-d\cdot P(|\widetilde{\omega}_j-\omega_j|> cN^{-\tau}) \nonumber\\
&\geq& 1-d\eta(1-N^{-2\tau}/\eta)^{m\lfloor n/k \rfloor},
\end{eqnarray*}
where $d$ is the cardinality of $\mathcal {M}$. The theorem is proved.
\hfill$\Box$

\bibliographystyle{asa}
\bibliography{papers}

\end{document}

%% file: self-define-HL.tex
\def\delete#1{\iffalse #1 \fi}

\def\bse{\begin{eqnarray*}}
\def\ese{\end{eqnarray*}}
\def\bee{\begin{enumerate}}
\def\eee{\end{enumerate}}
\def\bqe{\begin{eqnarray}}
\def\eqe{\end{eqnarray}}
\def\bed{\begin{description}}
\def\eed{\end{description}}
\def\bei{\begin{itemize}}
\def\eei{\end{itemize}}

\def\pmb#1{\setbox0=\hbox{#1}%
    \kern-.025em\copy0\kern-\wd0
    \kern.05em\copy0\kern-\wd0
    \kern-.025em\raise.0433em\box0 }
\def\pmbh#1#2{\setbox0=\hbox{#1}%
    \setbox1=\hbox{#2}%
    \kern-.025em\copy0\kern-\wd0
    \kern.05em\copy1\kern-\wd0
    \kern-.025em\raise.0433em\box0 }

\def\frac#1#2{{#1\over#2}}

\def\boxit#1{\vbox{\hrule\hbox{\vrule\kern6pt
   \vbox{\kern6pt#1\kern6pt}\kern6pt\vrule}\hrule}}

\def\listing#1{\vskip 4mm\begin{verbatim}\input#1 \vskip 4mm}
\def\thick#1{\hbox{\rlap{$#1$}\kern0.25pt\rlap{$#1$}\kern0.25pt$#1$}}

\def\pmbh{{\pmb h}}

\renewcommand\today{\ifcase\month\or
   Jan\or Feb\or Mar\or Apr\or May\or
   Jun\or Jul\or Aug\or Sep\or Oct\or Nov\or
   Dec\fi
   \space\number\day, \number\year}